\documentclass{egpubl}
\usepackage{eurovis2016}

\ShortPresentation      % uncomment for (final) Short Conference Presentation
 \electronicVersion % can be used both for the printed and electronic version

\ifpdf \usepackage[pdftex]{graphicx} \pdfcompresslevel=9
\else \usepackage[dvips]{graphicx} \fi

\PrintedOrElectronic

\usepackage{t1enc,dfadobe}

\usepackage{egweblnk}
\usepackage{cite}

\title{Visualization of Publication Impact}
\author[Eamonn Maguire, Javier Martin Montull, \& Gilles Louppe]
       {Eamonn Maguire$^{1}$, Javier Martin Montull$^{1}$, \& Gilles Louppe$^{2}$
        \\
         $^1$CERN, Geneva, Switzerland\\
         $^2$New York University, New York, USA
       }

\begin{document}

 \teaser{
  \includegraphics[width=0.97\linewidth]{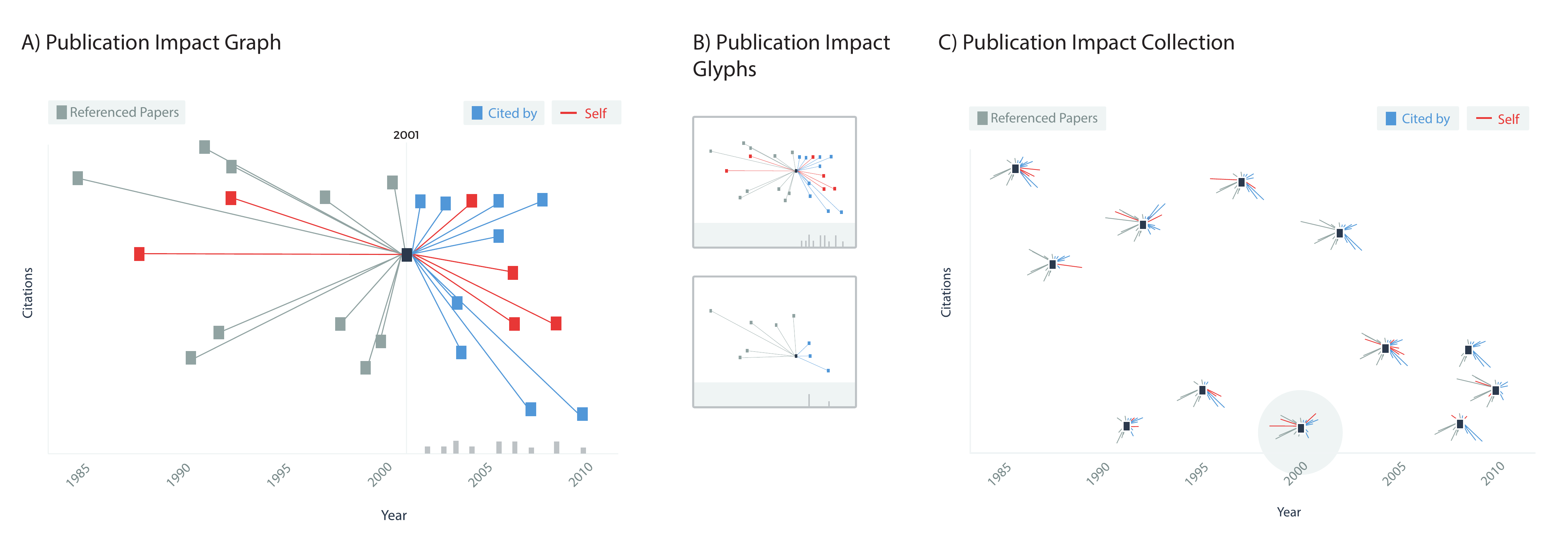}
  \centering
   \caption{A) A publication impact graph uses time on the X axis, and citations on the Y axis to visualise a publication in context with the references for that paper, and the papers that have cited it. B) Glyphs representations are a more compact version of A) with the citation histogram distinguish by its own section in the glyph. C) Visualise the publication space for an author, institution, or subject area by placing many impact glyphs in 2D space. Citation lines are truncated and expandable with interaction.}
 \label{fig:teaser}
 }

\maketitle

\begin{abstract}
Measuring scholarly impact has been a topic of much interest in recent years.
While many use the citation count as a primary indicator of a publications impact,
the quality and impact of those citations will vary.
Additionally, it is often difficult to see where a paper sits among other papers in the 
same research area.
Questions we wished to answer through this visualization were: is a publication cited less than publications in the field?;
is a publication cited by high or low impact publications?; and can we visually compare the impact of publications across a result set?
In this work we address the above questions through a new visualization of publication impact.
Our technique has been applied to the visualization of citation information in \textsc{InspireHep} (\url{www.inspirehep.net}), the largest high energy physics publication repository.
\end{abstract}

%-------------------------------------------------------------------------
\section{Introduction}

While a publications impact is currently viewed at the level of the number of citations, this metric can be misleading, with self-citations for instance
artificially driving up an articles perceived importance.
Moreover, the weight of a citation (how many times that paper has been cited) can vary but is not immediately available from any existing user interface or visualization tool.
Having a way to represent the impact of a publication, not only in the number of citations it received, but how important each of those citations was would provide an opportunity to assess a publications impact in context with related papers in the domain.

In this work we address the challenge of visualizing publication impact through a novel visualization that can exist in three states as shown in Fig. \ref{fig:teaser}:
as a standalone graph (see Fig. \ref{fig:teaser}A);
as a glyph design to provide overview level information about a publications impact (see Fig. \ref{fig:teaser}B); and
as an informative publication landscape composed of a set of the aforementioned glyphs \cite{borgo13} (see Fig. \ref{fig:teaser}C).

We utilize a comprehensive citation dataset from the largest curated high energy physics repository, \textsc{InspireHep}.
\textsc{InspireHep} maintains its own high quality citation engine to determine accurate citation counts for publications in high energy physics.

Our visualizations have been implemented in D3.js and are open source and available at \url{https://git.io/vayYO}.

The remainder of this paper will be organized in to five sections:
\textbf{related work} in Section \ref{sec:related_work} where we detail the most relevant related work;
\textbf{design} in Section \ref{sec:design} outlines the design processes involved in creating the visualizations in Fig. \ref{fig:teaser};
the \textbf{implementation} in Section \ref{sec:implementation};
\textbf{future work} in Section \ref{sec:future}; and
conclusions in Section \ref{sec:conclusion}.

\section{Related Work}
\label{sec:related_work}

Related work is sub divided in to: 1) visualization of citation networks; and 2) visualization of publication impact.

\subsection{Visualising Citation Networks}
As a natural fit to the data, network and more recently matrix-based techniques dominate the approaches used to visualize this type of data \cite{gibson13}.
Publications are generally represented as nodes in a graph which can be colored by their subject area or publication venue for example.
Directed edges indicate when a publication references another.
The citation count for a paper is computed through consideration of the number of incoming edges to a node.

\emph{CitNetExplorer} by van Eck and Waltman \cite{vanEck14} is an example of a citation visualization tool that utilized more graph-based approaches.
\emph{Citeology} by Matejka \emph{et al}\cite{matejka12} provides a context driven approach to visualizing a papers citations by arranging each reference and citation along the X axis by year of publication.
On the Y axis, each paper is represented by its title. On hovering over a publication title, all references and citations can be viewed as a pathway.
Additionally, Noel \emph{et al}\cite{noel03} devised a technique using minimal spanning trees to visualize co-citations and correlations between authors.

As is often the case with network visualization techniques, when the network becomes large, what is termed a `hairball` can form where nothing is visible anymore.   
Other techniques have been developed to navigate this issue.
CiteVis \cite{stasko13} for example uses a matrix to view papers. 
\emph{CiteRivers}\cite{heimerl16} is a powerful tool for the visual exploration of citation patterns that features the use of streams \cite{havre02}.

Finally, \emph{Hive Plots} \cite{krzywinski11} are a technique that could be used to reduce the visual complexity of large networks to make it easier to view within discipline/field citations (e.g. publications within high energy physics) or citations from external fields (e.g. citations from papers in high energy physics to mathematics).

\subsection{Visualising Publication Impact}
Publication impact is typically visualized by looking at the number of citations received by the paper.
Visualization tools such as \emph{Paperscape}\footnote{Paperscape \url{http://paperscape.org/}} use the area of a circle to represent publication importance.
Citation networks typically represent the impact of a publication by its connectedness in the graph.
An alternative, but effective representation is to plot a publication by its citation count (y axis) and time (x axis).
Altimetric\footnote{Altmetric \url{http://www.altmetric.com}}, a publication impact tracking service that considers social media shares, views, addition to citation management tools, and so on uses 2D plots to communicate the impact of a publication.

%-------------------------------------------------------------------------
\section{Design}
\label{sec:design}
There are numerous tools and techniques already available for the visualization of citation networks.
However, they focus primarily on authors or research subjects, and don't make it possible to compare the impact between publications.
The motivation of this work is based on \textsc{InspireHep} user requests to devise a solution that answers the following questions:
1) is a publication cited less than publications in the field?;
2) is a publication cited by high or low impact publications?; and
3) can we visually compare the impact of publications across a result set?

\begin{figure}[t!]
\centering
\includegraphics[width=.48\textwidth]{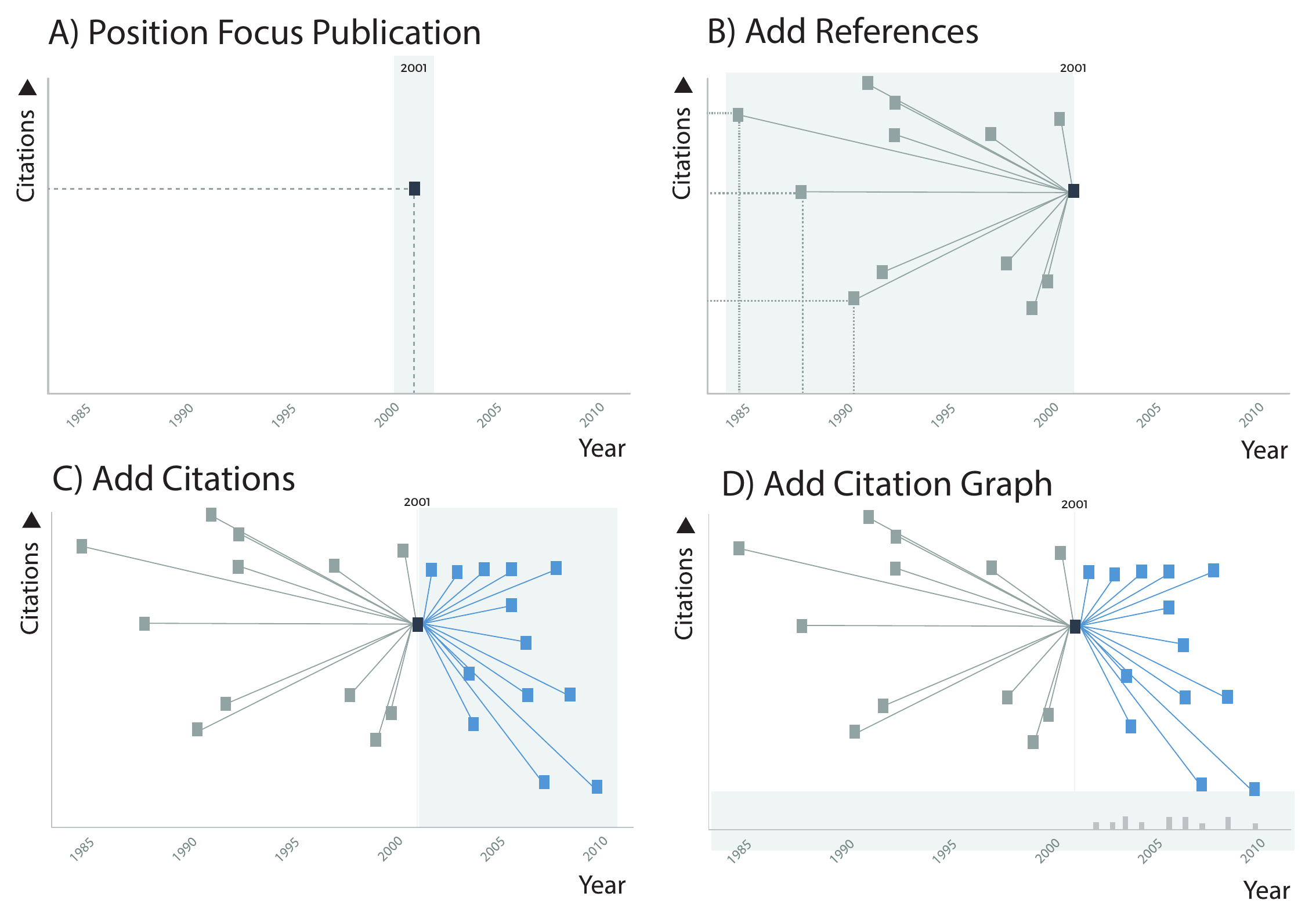}
\caption{A) First we position the publication of focus by its publication date and citation count.
B) We add each reference, again by its publication date and citation count and connect each node with an edge.
C) We repeat the process in B for citations.
D) A histogram is added to show the number of citations per year/month.}
\label{fig:graph_design}
\vspace{-10pt}
\end{figure}

The aim of our design is to take into consideration the questions posed in Section \ref{sec:design} to provide a visualization that can deliver important information to users across different resolutions.
The design takes the form of three interconnected parts:
1) impact graphs (detailed information for one paper);
2) impact glyphs (compact versions of the impact graph); and 
3) impact overviews (where we position many impact graphs for a subject area, author, institution, etc.).

\subsection{Impact Graphs}
\label{sec:impact_graph}

Given the questions the users wished to answer, and the information available, we started with the idea of an impact graph that can provide a way of visualizing a focus publication, its references, and citations. The impact graph is composed in the step wise way shown in Fig. \ref{fig:graph_design}.

\begin{figure}[ht!]
\centering
\includegraphics[width=.49\textwidth]{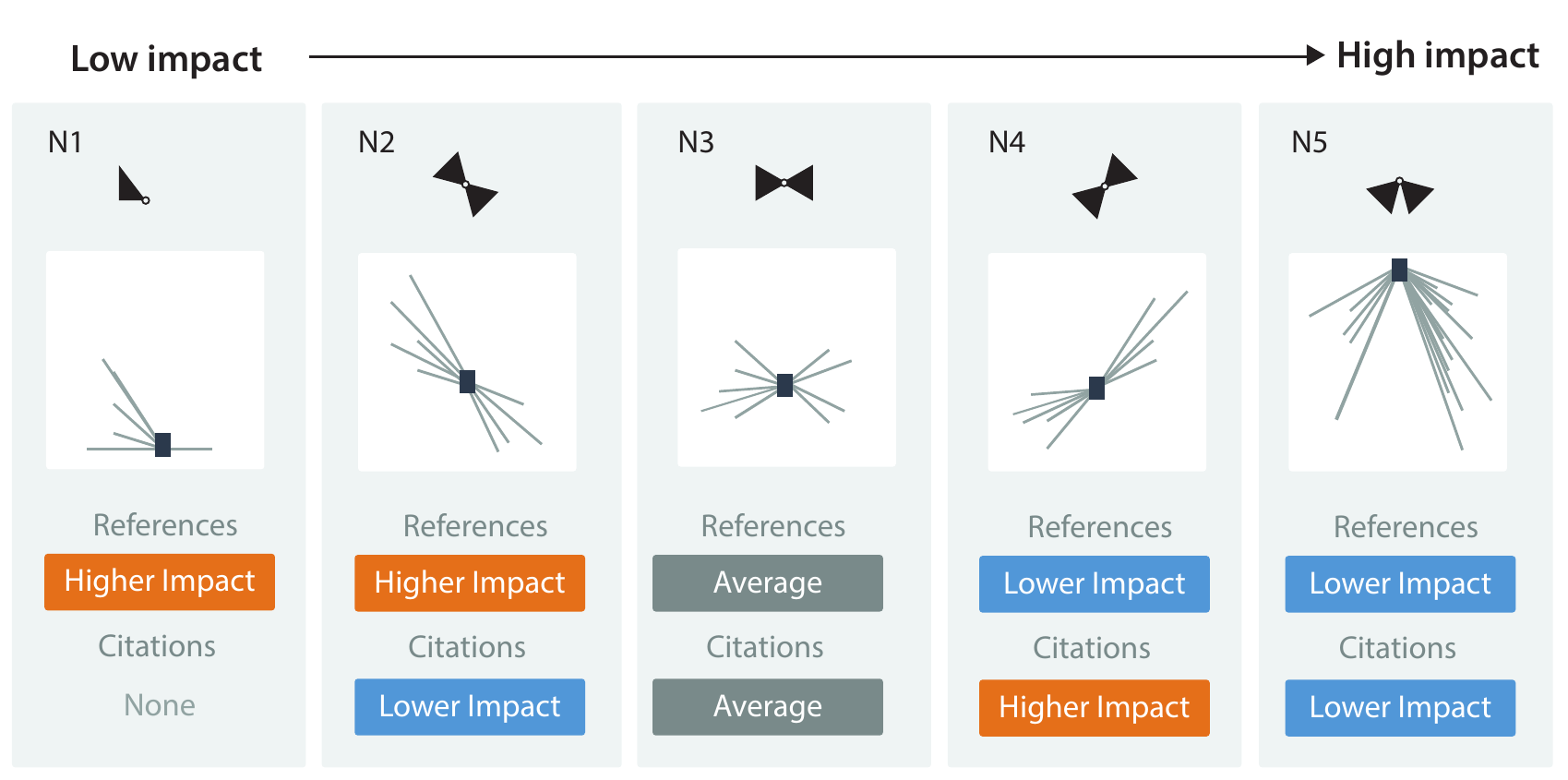}
\caption{Motifs showing general patterns that can be used to identify publications of varying impact.}
\label{fig:macro_structures}
\vspace{-10pt}
\end{figure}

Through the topological arrangement of a publications references and citations, it should be possible to define motifs, or frequently occurring patterns that correspond to publications of varying impact within their sphere of influence.
 
In Fig. \ref{fig:macro_structures} we show a selection of topologies defined from the observation of a large corpora of citation data.
We have identified five motifs that we believe adequately represent the common citation patterns for publications.
Each identified topological arrangement can point to papers of various levels of importance in their field depending largely on the citation counts of references and citations. 
For example, a publication may have a low number of citations, but the impact of that paper could be considered greater if those citing papers had high citation counts of their own (see Fig. \ref{fig:macro_structures} N4).

Conversely, a paper with a high number of citations may appear to be a high impact impact publication, however if all those citing papers have been cited less or if there are a large number of self citations, then the actual impact of the publication should be considered lower (see Fig. \ref{fig:macro_structures} N2). 

\subsection{Impact Glyphs}
\label{sec:impact_glyph}

\begin{figure}[t!]
\centering
\includegraphics[width=.48\textwidth]{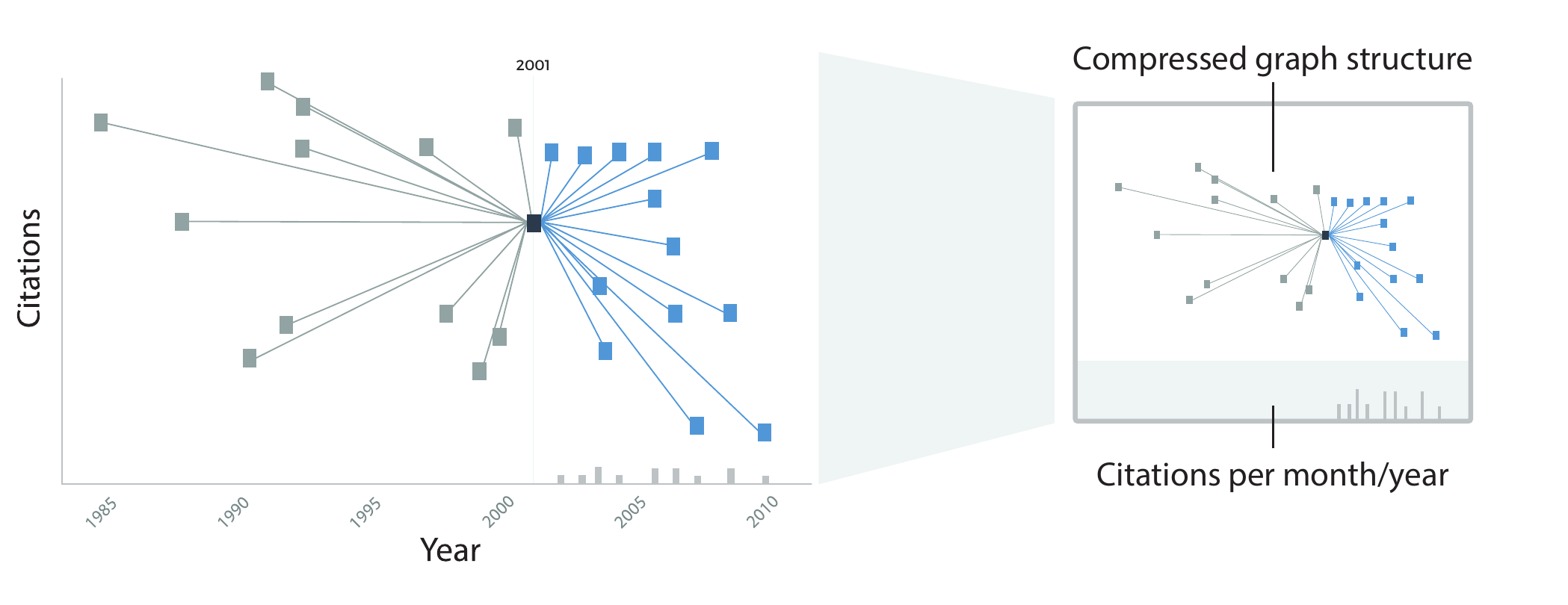}
\caption{Impact glyphs are impact graphs but without the axes.}
\label{fig:glyph_design}
\vspace{-10pt}
\end{figure}

As shown in Fig. \ref{fig:glyph_design}, \emph{Impact graphs} can be condensed in to \emph{impact glyphs} to show the general importance of a paper, number of citations and references (and their impact).
Their design considered the requirement to show the important features of an impact glyph even at low resolutions.

\begin{figure}[t!]
\centering
\includegraphics[scale=1]{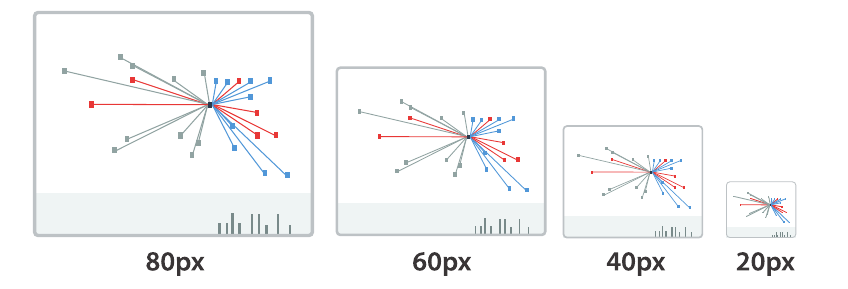}
\caption{Crush tests are a way of checking to ensure the key information is displayed even at low resolutions.}
\label{fig:glyph_crush}
\vspace{-10pt}
\end{figure}

We tested our design to ensure that key information such as the topological arrange of citations and references, citation density, and self-citations, could be seen at low resolutions.
Crush tests as introduced by Maguire \emph{et al}\cite{maguire12} allow for such comparisons in glyph designs.
Shown in Fig. \ref{fig:glyph_crush}, our glyph design has been subjected to crush tests from 80 pixels down to 20 pixel wide glyphs.
At 80 to 40 pixels, all important information is available.
Even high spatial frequency information such as that encoded in the citation graph is visible down to 40 pixels.
At 20 pixels, the topological arrangement is still evident showing a fairly average publication impact among the scope of related papers.

\subsection{Impact Overviews}
\label{sec:impact_overview}

\begin{figure}[ht!]
\centering
\includegraphics[width=.48\textwidth]{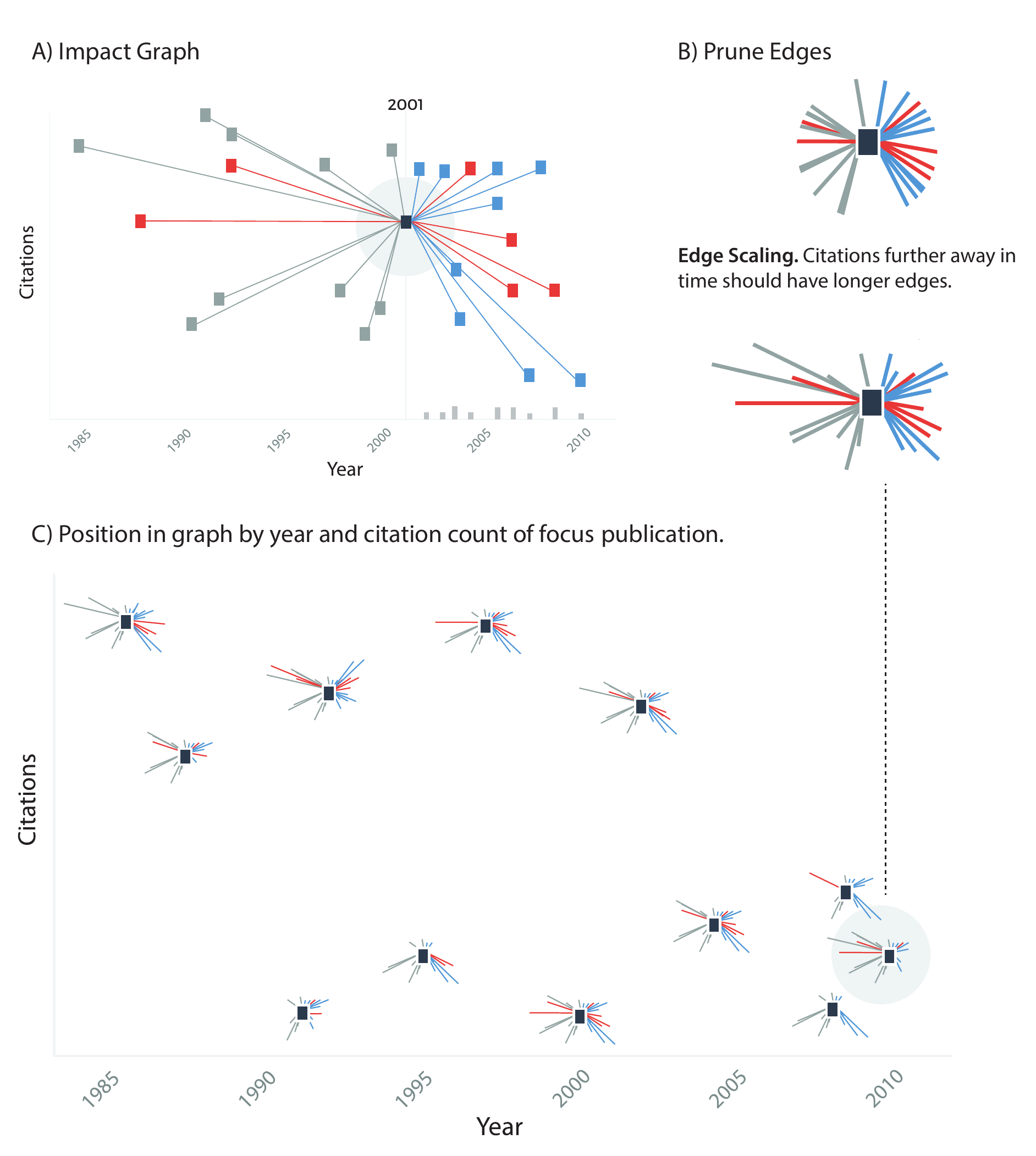}
\caption{A) We take a standard impact graph as a first step. B) Edges are truncated to avoid overlap with the edges of other publication items. Edge length is scaled to maintain the concept of publication date. C) Each glyph is positioned on a graph area spanning the minimum and maximum publication dates and citation counts.}
\label{fig:impact_overview}
\vspace{-15pt}
\end{figure}

Finally, impact overviews provide a way of viewing many papers from a subject area, author, collaboration, institution, etc. in a condensed view.
Illustrated in Fig. \ref{fig:impact_overview}, they take the core concepts from the impact graph and impact glyphs, but provide a way to layout the glyphs in 2D space in context with other publication impact glyphs.

These overviews are constructed through the use of a modified glyph that shows much the same information as for impact glyphs, however the edges aren't always drawn to the exact point in a graph that a reference or citation exists.
Instead, we draw a line that matches the citation count of the publication, however the time element is scaled in an attempt to avoid overlaps with other glyphs.
We are aware that with a large number of publications that there could be overcrowding.
To avoid this, we provide the option to change the transparency of glyphs so that the effect of overlaps is reduced.

\begin{figure*}[t!]
\centering
\includegraphics[width=.9\textwidth]{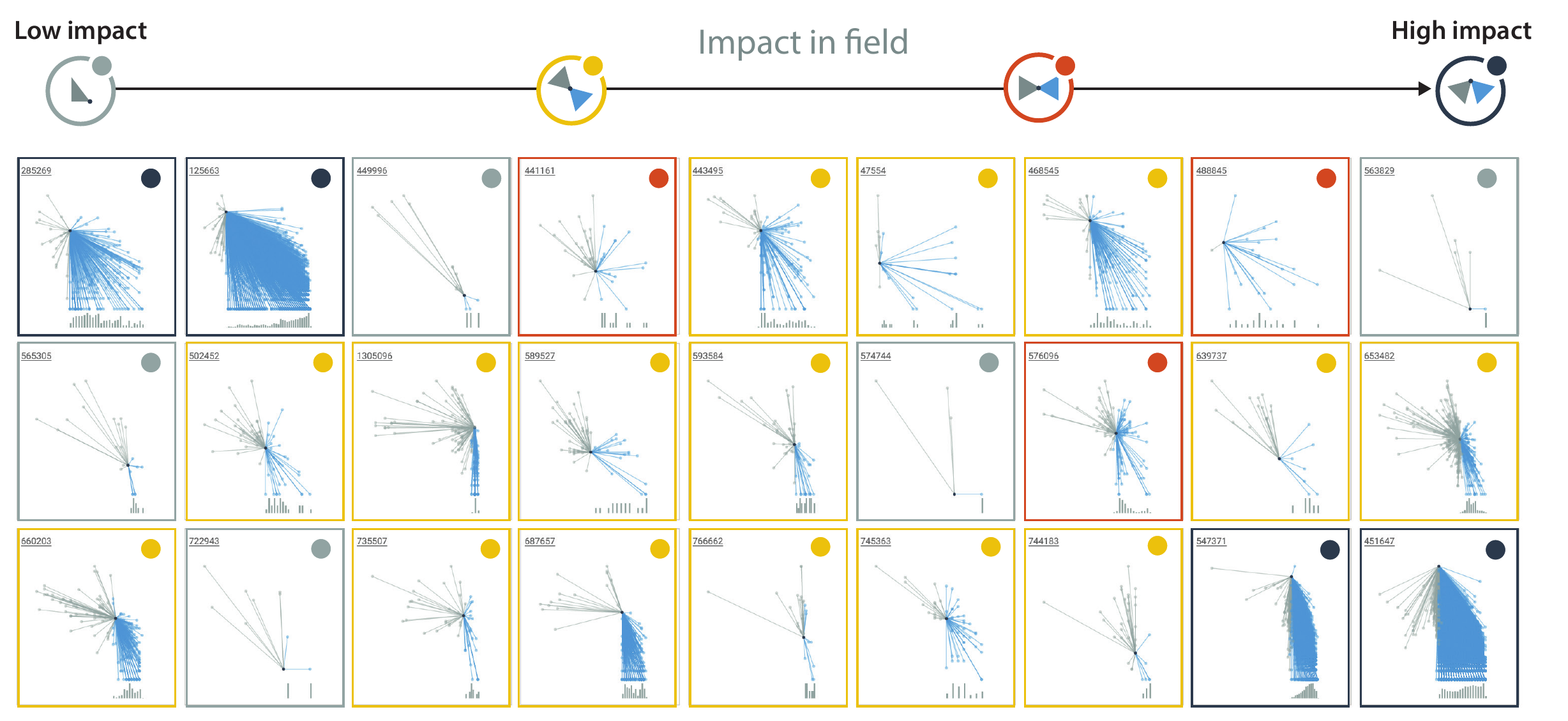}
\caption{Glyphs created for numerous \textsc{InspireHep} records show a number of impact motifs with low impact compared to their publication sphere, average impact, and high impact.}
\label{fig:glyph_matrix}
\vspace{-23pt}
\end{figure*}

%-------------------------------------------------------------------------
\section{Implementation}
\label{sec:implementation}

Our designs have been implemented in D3.js\cite{d3} and use a simple JSON data format.
All three modes of operation, to create publication impact graphs, impact glyphs, and impact overviews can be accessed from one library, and use the same overarching data format (multiple network definitions are consumed for impact overview visualizations).
The library is easily installable through \emph{bower} via the \emph{impact-graphs} package. 

Our publication impact graphs, glyphs, and overview visualizations are being added to the new \textsc{InspireHep} platform which will be released in the coming months where it will be used to visualize over 1.1 million publications.
We have run our visualization approach over many thousands of publications to produce output such as that shown in Fig. \ref{fig:glyph_matrix} where even in this this small subset, many of the motifs identified in Fig. \ref{fig:macro_structures} can be observed.

To exemplify the scalability of the approach, the glyph in the bottom right of Fig. \ref{fig:glyph_matrix} is visualizing publication 451647 from \textsc{InspireHep} (\emph{The Large N limit of superconformal field theories and supergravity}) which has over 11,000 citations.
Rendering speed is also important since we envisage these visualizations being optionally shown in search result pages.
With thousands of publications from \textsc{InspireHep}, we have observed rendering speeds of $<10ms$ for records with less than 500 citations.
For 11,000 citations, rendering takes $~400ms$.

Finally, our library comes with many options to enable configuration of: the Y scales from log to linear; the minimum and maximum citation counts and years (to facilitate easier between-glyph comparison); and automatic anomaly detection to highlight references made after or citations made before the publication date.
Such errors can point to issues with multiple versions of the same publication record.

\section{Future Work}
\label{sec:future}
With much of the functionality already present, future work will focus on an evaluation. Our visualizations have been designed with feedback from day to day users of \textsc{InspireHep}, however we do not assume that the encoding will be immediately familiar to all.
So far, our experience shows that users understand the visualization after a short introduction.
A full scale user evaluation will help to confirm this across the \textsc{InspireHep} user base.

\section{Conclusion}
\label{sec:conclusion}

We have presented a new glyph design for the visualization of publication impact. 
We have provided an implementation that can be immediately incorporated in to existing digital libraries for interactive use either as dedicated visualizations of a papers publication impact (impact graphs), as glyphs to accompany search results, or to be used as mass summarizations of publication impact across a database.

%-------------------------------------------------------------------------
\newpage
\bibliographystyle{eg-alpha-doi}
\bibliography{egbibsample}

%-------------------------------------------------------------------------

\end{document}